\documentclass[runningheads]{llncs}
\usepackage[utf8]{inputenc}
\usepackage{longtable}
\usepackage{supertabular,booktabs}
\usepackage{caption}
\usepackage{multirow}
\usepackage{graphicx}
\usepackage{float}
\usepackage{comment}
\usepackage{lscape}
\usepackage{graphics}
\usepackage{subcaption}
\usepackage{url}

\usepackage{breakurl}
\usepackage[breaklinks]{hyperref}
\begin{document}

\title{When Googling it doesn't work: The challenge of finding security advice for smart home devices
	\thanks{This is the author version of the article: Turner S., Nurse J., Li S. (2021) When Googling It Doesn’t Work: The Challenge of Finding Security Advice for Smart Home Devices. In: Furnell S., Clarke N. (eds) Human Aspects of Information Security and Assurance. HAISA 2021. IFIP Advances in Information and Communication Technology, vol 613. Springer, Cham. https://doi.org/10.1007/978-3-030-81111-2\_10}}

\titlerunning{Finding Security Advice for Smart Home Devices}

\author{Sarah Turner\orcidID{0000-0003-1246-1528} \and
Jason Nurse\orcidID{0000-0003-4118-1680} \and
Shujun Li}
\authorrunning{S. Turner et al.}

\institute{Institute of Cyber Security for Society (iCSS), University of Kent, Canterbury, UK\\ 
\email{\{slt41,j.r.c.nurse,s.j.li\}@kent.ac.uk}}

\maketitle     
\begin{abstract}
As users increasingly introduce Internet-connected devices into their homes, having access to accurate and relevant cyber security information is a fundamental means of ensuring safe use. Given the paucity of information provided with many devices at the time of purchase, this paper engages in a critical study of the type of advice that home Internet of Things (IoT) or smart device users might be presented with on the Internet to inform their cyber security practices. We base our research on an analysis of 427 web pages from 234 organisations that present information on security threats and relevant cyber security advice. The results show that users searching online for information are subject to an enormous range of advice and news from various sources with differing levels of credibility and relevance. With no clear explanation of how a user may assess the threats as they are pertinent to them, it becomes difficult to understand which pieces of advice would be the most effective in their situation. Recommendations are made to improve the clarity, consistency and availability of guidance from recognised sources to improve user access and understanding.

\keywords{Internet of Things, Cyber Security, Smart Home, Cyber Security Advice, Information, Online Search, Connected Home}
\end{abstract}

\section{Introduction}

Home Internet of Things (IoT) devices\footnote{The phrase ``home IoT devices'' here aligns with the list of devices found in \cite{dcms_code_2018}.} create different risks to their users than more traditional Internet-connected devices, such as personal computers. At an individual level, these include threats to physical safety, home security, personal control and privacy \cite{tanczer_emerging_2018}, and at a societal level, facilitating botnets and other Internet-based crime \cite{blythe_what_2020}. As these devices may come with little in-built security, these risks can quite quickly spread further than the individual device; as such the user must also understand how to manage the appropriate security of their entire home network. Home IoT devices are typically marketed on their minimal interfaces \cite{geeng_who_2019}, leaving the user to search elsewhere for guidance on issues such as cyber security. The availability of good quality, consistent and actionable information is crucial for keeping users safe and confident in their device use. Appropriately targeted levels of guidance for users is particularly necessary as cyber security is broadly considered a difficult topic for individuals to manage, despite there being a general acceptance of individual accountability for personal device use \cite{renaud_is_2018}. This is increasingly important for users, manufacturers, Internet Service Providers (ISPs) and policy-making bodies to understand and attempt to mitigate as sales in home IoT devices continue to grow apace, with users seemingly undeterred by frequent media stories of data breaches and other security risks.

This paper provides a review of cyber security information available on the Internet in relation to home IoT devices. It is driven by three primary research questions: what information is made available about cyber security threats posed to individuals using home IoT devices, what information is given around how to mitigate those threats, and what type of organisations or entities provide that information.  Using search methods that a typical user might undertake, we find that that advice that users are presented with is typically generalised and not sufficiently specific to act upon immediately, that the advice returned is often contradictory between sources, and that organisations that users would reasonably expect to have most responsibility for providing accurate content (manufacturers, governmental bodies, and ISPs) are not as prominently featured as they should be.

Following a brief literature review in Section~\ref{related_works}, and methodology in Section~\ref{sec:methodology}, we report our findings in Section~\ref{sec:findings}. Section~\ref{sec:discussion} considers the ways in which advice may need to be better tailored and managed to bolster users' understanding and willingness to act. Section~\ref{sec:limitations} considers limitations of the research, and how this could be addressed with future work.  Section~\ref{sec:conclusions} concludes the paper.

\section{Literature Review}
\label{related_works}

Previous research has looked at how users understand, evaluate and use cyber security methods. Cost, effort to set up and perceived inefficacy have been shown to stop individuals from adopting security tools such as anti-malware or password managers \cite{dupuis_use_2019}. Tabassum et al.~\cite{tabassum_i_2019} found that some home IoT device owners applied security knowledge learned from other contexts (such as from using computers and the Internet) when securing their home devices, despite the differences in threats posed and potential mitigating actions required.  It is widely recognised that this action, in part, arises from a wide-spread lack of accurate mental models about these devices \cite{zeng_end_2017}, unless the user is already very technically minded \cite{luger_like_2016}.

Even if individuals do act to implement cyber security measures at home, they could be overwhelmed with the number of actions that are deemed to be essential: Redmiles et al.~\cite{redmiles_comprehensive_2020} found 374 pieces of actionable advice in reviewing publicly available documentation, and argued that what is needed is effective prioritisation of that advice. Prior to purchase, users rarely look for security and privacy information, but note that it is impossible to find if they do \cite{emami-naeini_exploring_2019}. Gcaza argued that security awareness is a necessary requirement for communities to consider themselves ``smart'' \cite{gcaza_cybersecurity_2018}: enhanced levels of clarity have been called for in both governmental and manufacturer's advice, to promote tangible steps to security \cite{van_steen_what_2020}, a better understanding of how the technology works \cite{voit_not_2020}, and how the user is affected in the case of a breach \cite{zou_you_2019}. This clarity should extend to the practices of the manufacturer, in particular in relation to privacy and security concerns \cite{kulyk_does_2020}. There is a clear benefit to this: users will pay a premium for devices that have prominent details about security features \cite{blythe_security_2019}.

\section{Methodology} 
\label{sec:methodology}

In order to understand what a home IoT device user might encounter when searching for information about how to secure devices that they may have, the decision was taken to search the Internet for cyber security guidance. This was done both in relation to general devices, using general search terms and reviewing the results that mentioned home IoT devices specifically, as well as for the most popular devices in the UK at this time: smart TVs (and streaming devices), and smart home assistants \cite{techuk_state_2020}. This decision was made because of the proportion of individuals voluntarily using these devices; findings for these specific types of devices may offer more value by virtue of their ubiquity than other device types. Recent research has used similar practices in relation to posted user reviews \cite{oygur_raising_2020} to understand what type of information users may encounter online on specific topics. 

\begin{table}[ht]
\caption{Generalised search queries}
\label{tab:search_queries}
\centering

\begin{tabular}{@{}p{0.4\linewidth}p{0.5\linewidth}@{}}
\toprule
\multicolumn{2}{c}{\textbf{Search terms}} \\ \midrule
Cyber security information & Cyber security charities \\
Cyber security awareness & Internet of Things cyber security help \\
Cyber security knowledge & Cyber security help \\
Cyber security education & Cyber security support \\
Cyber security learning & Smart devices cyber security help \\
Cyber security training & How to stop being hacked  \\
Cyber security organisations & How to secure my devices \\
 \bottomrule
\end{tabular}
\end{table}

The general device resources were sought through search terms listed in Table~\ref{tab:search_queries}, and reflect a final list after researcher experimentation with various similar terms. Using these general search terms, pages in the results that had references to home IoT devices were captured for analysis.  Search terms relating to specific devices took the form ``How to secure my smart TV/streaming device/smart speaker''  along with ``[manufacturer name] [device name] security'' (e.g., ``Amazon Echo security''). Specific brands were chosen based upon lists of ``Top devices for 2020'' focused on UK consumers.\footnote{\url{https://www.techradar.com/uk/news/best-smart-speakers}, \url{https://www.techradar.com/uk/news/best-tv}, \url{https://www.techadvisor.co.uk/test-centre/digital-home/best-media-streaming-box-3580569/}} Having logged out of all browser accounts, cleared user history and using a VPN connection to a different IP address in the UK, the search terms were entered into three search engines: Google, Bing and Duck Duck Go,\footnote{These account for nearly 97\% of all UK search engine traffic as of July 2020 \cite{johnson_market_2020}.} and non-paid search results from the first two pages of each search query were captured, on the understanding that less than one in ten users are likely to go to the second page of search results \cite{ray_we_2019}. The pages were retrieved between August and December 2020. For both results from the generalised search and specific devices searches, each page was then reviewed, and those that had content referring to home IoT devices were then taken forward for analysis. Following methodology from \cite{blythe_what_2020} and \cite{turner_the_2020}, a number of predefined criteria were captured from each page, including who produced the information and when, the type of devices considered, and the threats and advice given.

\section{Results} 
\label{sec:findings}

\subsection{Sources of information}
\label{sec:sources}

The prominence of news and opinion outlets is clear in the results. 125 sources (53.41\%) of the 234 organisations with web pages considered in the review were either recognised news organisations (such as The Guardian, Wired, CNet) or websites offering news and opinion pieces of varying levels of specialty and expertise, ranging from personal blogs to user-facing technology sites (such as PC Mag, ZD Net). The search also returned a volunteer-run cyber security helpline,\footnote{\url{https://www.thecyberhelpline.com}} offering help across a wide range of cyber security issues. We also found that the favourable rankings of more traditional news sites acted to suppress sources of advice and information about device security in favour of prior security and data breaches: notably, a 2014 breach relating to Philips' smart TV range still dominated the first two pages of results, even in Google's Featured Snippets,\footnote{For more on Google's Featured Snippets, see \url{https://support.google.com/websearch/answer/9351707}.} despite the age of the story. Although the majority of individual web pages returned were dated 2019 and 2020 (228 web pages, 53.40\%, of 427 total web pages), 91 were undated, and 2 websites (from a retailer, and anti-malware provider) had content dating from 2011 (from the date given in the body of the article).

Only nine information sources of the 234 organisations were affiliated with global governmental departments; there were three consumer protection bodies (such as Which? and Consumer Reports) and five additional not-for-profit or charitable bodies. Conversely, bodies that may have been trying to sell a service related to security were much more common: there were nine anti-malware providers (such as Malwarebytes and Kaspersky), and firms offering cyber security services (such as BullGuard, Digital Guardian and Cytelligence) accounted for 21 pages. There were 13 forum sites, both third-party (Reddit, Stack Exchange) and manufacturer community pages. There were no sites from ISPs returned in the results.

\begin{table}[ht]
\caption{Advice and Threat Types}
\begin{subtable}[]{0.45\linewidth}
\centering
\caption{Top five: threat types}
\label{tab:threats}
\begin{tabular}{@{}ll@{}}
\toprule
\multicolumn{1}{c}{\textbf{Type of threat}} & \multicolumn{1}{c}{\textbf{Count}} \\ \midrule
Unauthorised access & 144 \\
Malware & 22 \\
Data theft & 13 \\
Botnet & 9 \\
Ransomware & 8 \\
 \bottomrule
\end{tabular}
\end{subtable}
\begin{subtable}{0.45\linewidth}
\centering
\caption{Top five: advice types}
\label{tab:advice}
\begin{tabular}{@{}ll@{}}
\toprule
\multicolumn{1}{c}{\textbf{Type of advice}} & \multicolumn{1}{c}{\textbf{Count}} \\ \midrule
Strong password management & 149 \\
Limit data access & 145 \\
Better home network security & 143 \\
Turn off features/devices & 117 \\
Update software  & 113 \\
 \bottomrule
\end{tabular}
\end{subtable}
\end{table}

\subsection{Reported threats}
\label{sec:reported_threats}

Discussions about cyber security typically arise from the need to secure something from a specific and meaningful threat. In the review, 57 individual types of threats were raised; for the top five, see Table~\ref{tab:threats}. 144 websites referred to some form of unauthorised access to devices, most typically ``hacking'', without further explanation (Table~\ref{tab:advice_table}, \#1). 39 web pages focused on either how to manage after you have been hacked or avoiding being hacked, typically presenting reactive advice rather than explaining why it may be necessary to take proactive measures ahead of an event (Table~\ref{tab:advice_table}, \#2). 
Malware and ransomware were mentioned a total of 30 times, with theft of personal data being mentioned 13 times. Botnets were referenced nine times. It is noticeable how many types of threat were referenced only once or twice throughout the review. 26 types of threat came up only once (examples ranging from domestic abuse, to ghostware and hacktivism). Lack of personal knowledge was framed as a threat (rather than a potential vulnerability) in five instances (Table~\ref{tab:advice_table}, \#3). In some cases, the publication of specific academic or industry reports were reflected in the reporting of several news sources (Table~\ref{tab:advice_table}, \#4). In these cases, the threats reported upon are typically accompanied by the researchers' views on how to mitigate the risk, albeit at a high level, often without accompanying links to manufacturer guidance for specific devices.

\subsection{Types of advice needed and provided}
\label{sec:advice_needed_provided}

In total, there were 1,342 pieces of advice counted in the reviewed web pages, which, when coded for advice type provided a total of 54 unique topics. The top five advice types are listed in Table~\ref{tab:advice}.

\begin{landscape}

\begin{table}[ht]
\caption{Examples of advice given (as referenced throughout text)}
\label{tab:advice_table}
\centering
\scalebox{0.57}{
\begin{tabular}{@{}llll@{}}
\toprule
\textbf{Reference} & \textbf{Source/date} & \textbf{Issue raised} & \textbf{Link (all last accessed 31 March 2021)}\\ \midrule
1 & \begin{tabular}[c]{@{}l@{}}IoT for All (2020)\\ (IoT blog)\end{tabular} & \begin{tabular}[c]{@{}l@{}}Generic threat explanation:``...they leave us vulnerable to cyber crime...\\ {[}IoT devices{]} are top targets for hackers.''\end{tabular} & \url{https://www.iotforall.com/iot-cyber-security-2}\\
2                  & \begin{tabular}[c]{@{}l@{}}Lifewire (2019)\\ (consumer technology blog)\end{tabular}                   & \begin{tabular}[c]{@{}l@{}}Reactive security guidance - things to do after you have been ``hacked'': ``no matter \\ how you were hacked, you're feeling vulnerable.''\end{tabular}                                                     & \begin{tabular}[c]{@{}l@{}}https://www.lifewire.com/securing-your-home\\ -network-and-pc-after-a-hack-2487231\end{tabular}                                   \\
3                  & \begin{tabular}[c]{@{}l@{}}IoT Wiki (2019)\\ (IoT enthusiast blog)\end{tabular}                 & \begin{tabular}[c]{@{}l@{}}Lack of knowledge a threat: ``many individual users...still lack information about the risk'' '\end{tabular}                                                                                       & \begin{tabular}[c]{@{}l@{}}https://internetofthingswiki.com/biggest-\\security-issues-iot-devices-face/1344/\end{tabular}                               \\
4                  & Tech Crunch (2019)                                                                                     & Report of FBI advice on smart TV security                                                                                                                                                                                              & \begin{tabular}[c]{@{}l@{}}https://techcrunch.com/2019/12/01/\\ fbi-smart-tv-security/\end{tabular}                                                          \\
5                  & \begin{tabular}[c]{@{}l@{}}Now TV (undated)\\ (streaming devices)\end{tabular}                         & Password guidance:``DON'T use a word that's found in the dictionary''                                                                                                                                                                  & \begin{tabular}[c]{@{}l@{}}https://help.nowtv.com/article/tips-to-help-you-\\ keep-your-account-secure\end{tabular}                                          \\
6                  & \begin{tabular}[c]{@{}l@{}}Comparitech (2020)\\ (consumer technology blog)\end{tabular}                & \begin{tabular}[c]{@{}l@{}}Password guidance: "Make changing the router password \\ part of your monthly routine"\end{tabular}                                                                                                         & \begin{tabular}[c]{@{}l@{}}https://www.comparitech.com/blog/information-\\ security/secure-home-wireless-network/\end{tabular}                               \\
7                  & \begin{tabular}[c]{@{}l@{}}CSO Online (2016)\\ (technology risk news site)\end{tabular}                & \begin{tabular}[c]{@{}l@{}}Don't use devices as intended: ``Don't connect your devices unless you need to...\\ turn off UPnP...be wary of cloud services''\end{tabular}                                                                & \begin{tabular}[c]{@{}l@{}}https://www.csoonline.com/article/3085607/8-\\ tips-to-secure-those-iot-devices.html\end{tabular}                                 \\
8                  & \begin{tabular}[c]{@{}l@{}}Lifehacker (2018)\\ (consumer blog)\end{tabular}                            & \begin{tabular}[c]{@{}l@{}}Non-specific advice: "If you're lucky, your router can broadcast a \\ `guest network'...''\end{tabular}                                                                                                     & \begin{tabular}[c]{@{}l@{}}https://lifehacker.com/how-to-keep-your-friends\\ -from-trolling-your-chromecast-1828805478\end{tabular}                          \\
9                  & \begin{tabular}[c]{@{}l@{}}Cytelligence (undated)\\ (cybersecurity service)\end{tabular}               & \begin{tabular}[c]{@{}l@{}}Non-specific advice: 10 ten list with no further details \\ (e.g.``Stick with protected devices only...Disable unnecessary features...Secure your\\ network fully'')\end{tabular}                           & \begin{tabular}[c]{@{}l@{}}https://cytelligence.com/cyber-security-and-\\ smart-devices/\end{tabular}                                                        \\
10                 & \begin{tabular}[c]{@{}l@{}}Digital Trends (2021)\\ (consumer technology blog)\end{tabular}             & How to secure your Alexa device (with 12 suggestions)                                                                                                                                                                                  & \begin{tabular}[c]{@{}l@{}}https://www.digitaltrends.com/home/how-to\\ -secure-your-alexa-device/\end{tabular}                                               \\
11                 & \begin{tabular}[c]{@{}l@{}}Wired (2020)\\ (technology magazine)\end{tabular}                           & \begin{tabular}[c]{@{}l@{}}Guest networks: ``grant your you guests access to a Wi-Fi connection \\ without letting them get at the rest of your network --- \\ your Sonos speakers, the shared folders on your laptop..."\end{tabular} & \begin{tabular}[c]{@{}l@{}}https://www.wired.com/story/secure-your-\\ wi-fi-router/\end{tabular}                                                             \\
12                 & \begin{tabular}[c]{@{}l@{}}Kaspersky (undated)\\ (anti-malware software)\end{tabular}                  & Guest networks: ``{[}set{]} up guest networks for your IoT home devices''                                                                                                                                                              & \begin{tabular}[c]{@{}l@{}}https://www.kaspersky.com/resource-center/\\ threats/how-safe-is-your-smart-home\end{tabular}                                     \\
13                 & \begin{tabular}[c]{@{}l@{}}How-To Geek (2020)\\ (consumer technology blog)\end{tabular}                & \begin{tabular}[c]{@{}l@{}}Guest networks: ``you would connect all your IoT devices...\\ and actual guests to the guest network''\end{tabular}                                                                                         & \begin{tabular}[c]{@{}l@{}}https://www.howtogeek.com/659084/\\ how-secure-is-your-home-wi-fi/\end{tabular}                                                   \\
14                 & \begin{tabular}[c]{@{}l@{}}Google (undated)\\ (Android devices)\end{tabular}                           & How to log your child into their Android device                                                                                                                                                                                        & \begin{tabular}[c]{@{}l@{}}https://support.google.com/families/answer/\\ 7158477?hl=en\end{tabular}                                                          \\
15                 & \begin{tabular}[c]{@{}l@{}}Help Cloud (undated)\\ (consumer security service)\end{tabular}             & Buy a more secure router: ``{[}invest{]} in a sound WiFi router''                                                                                                                                                                      & \begin{tabular}[c]{@{}l@{}}https://www.helpcloud.com/blog/cybersecurity\\ -experts-and-iot-smart-devices-and-smart-homes/\end{tabular}                       \\
16                 & \begin{tabular}[c]{@{}l@{}}Ready.gov (undated)\\ (US governmental resource)\end{tabular}               & \begin{tabular}[c]{@{}l@{}}Implication of need to buy more security: ``{[}use{]} a password manager...use \\ antivirus solutions...use a VPN...''\end{tabular}                                                                         & https://www.ready.gov/cybersecurity                                                                                                                          \\
17                 & \begin{tabular}[c]{@{}l@{}}PCWorld (2019)\\ (technology news site)\end{tabular}                        & \begin{tabular}[c]{@{}l@{}}Implication of need to buy more security: ``Our favourite password manager is \\ xxxx...you'll need to pay an annual fee, but it's worth it.''\end{tabular}                                                 & \begin{tabular}[c]{@{}l@{}}https://www.pcworld.com/article/3332211/\\ secure-android-phone.html\end{tabular}                                                 \\
18                 & \begin{tabular}[c]{@{}l@{}}Real Simple (2020)\\ (consumer blog)\end{tabular}                           & \begin{tabular}[c]{@{}l@{}}Buy reputable devices: ``If you want to have IoT devices around...a wiser\\ route is going to be by shopping in Apple or Google's walled gardens...''\end{tabular}                                          & \begin{tabular}[c]{@{}l@{}}https://www.realsimple.com/work-life/\\ technology/safety-family/smart-home\\ -cyber-security\end{tabular}                        \\
19                 & \begin{tabular}[c]{@{}l@{}}Norton (undated)\\ (anti-malware software)\end{tabular}                     & \begin{tabular}[c]{@{}l@{}}Choose based on privacy and data policies: ``What are the privacy policies? \\ Will the provider store your data or sell it to a third party? How are updates enabled?''\end{tabular}                       & \begin{tabular}[c]{@{}l@{}}https://us.norton.com/internetsecurity-iot-smart\\ -home-security-core.html\end{tabular}                                          \\
20                 & \begin{tabular}[c]{@{}l@{}}The Guardian (2020)\\ (news site)\end{tabular}                              & Coverage of Sonos' decision to stop software updates for old devices                                                                                                                                                                   & \begin{tabular}[c]{@{}l@{}}https://www.theguardian.com/technology/2020/\\ jan/23/sonos-to-deny-software-updates-to-owners\\ -of-older-equipment\end{tabular} \\
21                 & \begin{tabular}[c]{@{}l@{}}eBuyer (2018)\\ (consumer technology blog)\end{tabular}                     & \begin{tabular}[c]{@{}l@{}}Software updates:``You should always update your smart devices...as soon as \\ it becomes available''\end{tabular}                                                                                          & \begin{tabular}[c]{@{}l@{}}https://www.ebuyer.com/blog/2018/10/smart\\ -devices-and-security/\end{tabular}                                                   \\
22                 & \begin{tabular}[c]{@{}l@{}}National Cyber Security\\ Centre (2019)\\ (UK government body)\end{tabular} & Wiping device of data: ``you should first perform a factory reset.''                                                                                                                                                                   & \begin{tabular}[c]{@{}l@{}}https://www.ncsc.gov.uk/guidance/smart-devices\\ -in-the-home\end{tabular}                                                        \\ \bottomrule
\end{tabular}}
\end{table}

\end{landscape}

There were 149 separate instances of recommended strong password management (11.10\% of the total pieces advice given), many of which gave advice contrary to the current guidance from the UK's National Cyber Security Centre (NCSC) to use three random words to create a strong password. For example, two manufacturers explicitly suggested that words found in the dictionary should not be used (Table~\ref{tab:advice_table}, \#5), and suggestions to change passwords frequently were also common (Table~\ref{tab:advice_table}, \#6).

Limiting the access services have to personal data was the second most frequent type of advice given in the reviewed web pages (145 instances; 10.80\%), although precise guidance as to what this means for specific devices was not generally explained. Disabling some features (such as Universal Plug and Play) or turning off the device (or router, or WiFi) altogether was the fourth most common (117; 8.71\%). The trade-offs of doing these actions were again, largely unexplored (Table~\ref{tab:advice_table}, \#7). Specificity of advice was a common problem --- the heterogeneity of devices left some pages assuming that devices had particular functionality as the premise of their advice (Table~\ref{tab:advice_table}, \#8), or providing a list of things to do with no guidance at all (Table~\ref{tab:advice_table}, \#9). Other pages gave so much advice as to run the risk of seeming overwhelming (Table~\ref{tab:advice_table}, \#10).

There were 143 instances of advice around improving the strength of home networks. Advice around improving the strength of the user's home network is particularly difficult to follow, as the exact, typically relatively technical, steps vary upon the router in the house. In the general searches returned, there was no guidance about smart home security provided by ISPs. Without further searching in relation to the router owned by the individual, at first glance it is impossible for the reader to know which pieces of advice (such as ``use a VPN'' or ``set up a guest network'') would be feasible for their current router.  Setting up a guest network, in particular, was recommended, but the specifics of doing so were varied: some pages suggested putting all the user's devices on one network and anyone external on the other (Table~\ref{tab:advice_table}, \#11); others suggested keeping home IoT devices on one network, and the users' other devices and guests on the second (Table~\ref{tab:advice_table}, \#12); and there was also suggestion to keep your personal non-IoT devices on one, and your home IoT devices and guests on the other (Table~\ref{tab:advice_table}, \#13).

When manufacturer's pages were returned in the reviewed web pages they were typically in a wiki-format, for a very specific topic --- focusing how to change a specific setting rather than why you might do this --- with minimal visual guidance: a checklist of steps to perform a specific activity on a specific device (Table~\ref{tab:advice_table}, \#14). In contrast, sites not affiliated with manufacturers offer more generic advice. Not only did they provide little to specific device guidance or explanation as to what that would protect against, but they frequently suggested additional products that come at additional costs. Some are explicit: buying a more secure router (Table~\ref{tab:advice_table}, \#15), or, less clearly, products and services that can come with a cost, such as anti-malware, VPNs or password managers (Table~\ref{tab:advice_table}, \#16, \#17). Other advice given includes to be choosy with home IoT device providers (even at a risk of becoming locked into a single provider) (Table~\ref{tab:advice_table}, \#18), and performing pre-purchase checks such as reading privacy and data sharing/selling policies (Table~\ref{tab:advice_table}, \#19).

There was a striking lack of information about end of life device management, with the exception of the negative press relating to Sonos' decision to stop supporting older models in early 2020 (Table~\ref{tab:advice_table}, \#20), and general advice to ``update software'' (but not explicitly to be aware of the end of the supported life of your device) (Table~\ref{tab:advice_table}, \#21). Only the NCSC discussed wiping a device at the point of reselling or throwing away (Table~\ref{tab:advice_table}, \#22).

\section{Discussion} 
\label{sec:discussion}

A significant proportion of the guidance discovered in the reviewed web pages was not actionable for home IoT devices without further understanding or learning by the reader. The heterogeneity of home IoT devices, and the situations in which they are used, means that there there may be best practices that are specific to the device and its use. Different designs mean that users cannot guarantee that they will be able to follow steps to disable settings, for example, to adhere to best security practices, assuming the specific device they have has the functionality to allow the user to access and alter security settings.  Different threats mean that some users may be best off following different advice for the same device, but without an ability to accurately assess the threats and risks that the device poses to them, users are likely to fall back to behaviours that have worked for them before, which may not be appropriate in this case \cite{tabassum_i_2019}.

Furthermore, the most appropriate point to modify security settings may be at the home network, and not device, level. Calls to alter router settings, for example, are assuming that users have the technical confidence, sufficient access to the controls within the home setting, and that their routers have the functionality to do so, none of which may be the case \cite{zeng_end_2017}. Additional suggestions to use more software --- ideally, purchase software --- is problematic: it introduces another barrier to effective cyber security for those who cannot afford it, and it is unclear how to apply such software across all devices in the home, if it is even possible to do so. The attrition rates for use of such software is likely to be high, particularly if its value in protecting devices is not visible or obvious \cite{dupuis_use_2019}.

Governmental and consumer awareness resources did appear, often low down in results. Despite their relative trustworthiness and validity as relevant and impartial cyber security information, such resources are often indistinguishable from other sources in search results. These other sources may have financial interests in the framing of their advice (such as anti-malware providers), or the guidance may be from irrelevant or out of date sources. Users would benefit from higher placement in search results of official guidance from governmental agencies and manufacturers to try and provide up to date, specific information; inspiration could be taken from the work done to place prominent information from recognised expert bodies at the top of search results relating to COVID-19.

Advice to choose devices based upon more agreeable privacy policies or calls to do research before purchase and buy ``more secure devices'' highlight a lack of congruence between the advice and real life. Privacy policies are notoriously hard to read and comprehend \cite{renaud_how_2018}, and offer no ability for the user to negotiate the terms of their use. Calling upon users to research devices prior to purchase suggests that sufficient information is available to make a useful comparison of security features --- not only is it hard to find this information, it may not be meaningful or useful when found \cite{emami-naeini_exploring_2019}.

Providing standardised labels on packaging to provide information on fundamental security features may be helpful to help users determine what is important to them at the time of purchase\cite{emami_naeini_ask_2020}, however manufacturers need to help users to assess and review their security settings throughout the life of their devices. This could take the form of periodic notifications on the device or associated app, reminding users to check key risk areas for a given device. This, of course, may be device specific, but manufacturers could use the opportunity to target common areas of concern based upon market intelligence or user research to ensure that users are given an opportunity to secure the most pressing risks. Manufacturers should avoid confusion by only providing guidance that is in line with the regional governmental cyber security agency, or in the case of international manufacturers, picking advice from respected agencies or bodies, and referencing and linking back to those bodies so that users can see the underlying guidance themselves. As the results of the website review show, conflicting advice is abundant as a result of the number of expert opinions in the field, and so manufacturers can help users understand why they are promoting the security practices that they are. This also provides users an opportunity to learn about the evolving nature of cyber security information, and promotes the need for periodic reviews of the user's security setup. Making users aware that guidance is dynamically evolving, and explaining how they will receive updated advice, is crucial, and facilitates user learning. 

Before being able to manage risks effectively, however, users need to have more meaningful guidance about the types of threats that their devices may pose, so that they can appropriately evaluate what risk management means to them. This is a complex area, given the potential for misuse, abuse, and power imbalances \cite{ehrenberg_the_2021}. However, manufacturers of devices could produce and point users to common device use cases, for example, with different permutations of household device use (including how children and visitors may use the device). These use cases could explain the potential threats to the device in the situation, the implications of those threats, and how to mitigate those risks based upon the security features of the device.  This would also be beneficial for ISPs to offer their customers in relation to home router setup, to ensure insecure devices do not pose unexpected threats, both inside and outside of the home.

\section{Limitations and Future Work}
\label{sec:limitations}
This work was an exploratory piece of research, to determine what the Internet offered users when a number of generalised queries, and searches based upon the most popular devices reported in a recent survey were undertaken. The queries were researcher-generated, meaning that they may not exactly reflect the types of queries typical home users would perform. Decisions to limit the pages used in the search may also not reflect a user's behaviour when looking for a specific answer. While we did our best to mirror a reasonable keyword selection process and user-oriented approach to pages viewed, future work should involve users to generate these search terms, and use a more precise understanding of when users might stop looking for answers on a page. It may also be useful to do a wider review, as limiting the research to a handful of specific devices may ignore advice that is necessary for the security of other types of devices. The search results also point to the complex role that routers have in the smart home. Repeating the work with routers included as a specifically searched-for device may be beneficial.

\section{Conclusions} 
\label{sec:conclusions}
Through a review of web pages, this research has shown that finding reputable, actionable and coherent guidance on how to approach securing home IoT device against cyber security threats is challenging. Users are confronted by an overwhelming number of resources, often with little direct credibility or specific actionable advice. We consider that improvements could be made by device manufacturers in particular in creating clearer, more actionable content, as well as a need for search engine results to reflect more prominently those resources from relevant organisations (notably manufacturers and governmental bodies) to ensure users find the most specific advice for their situation.

\bibliographystyle{splncs04}
\bibliography{main}

\end{document}